\documentstyle{mn}
\ifnfsstwo

\fi

\ifnfssone
  \newmathalphabet{\mathit}
    \addtoversion{normal}{\mathit}{cmr}{m}{it}
    \addtoversion{bold}{\mathit}{cmr}{bx}{it}

\fi

\ifoldfss

\fi

\loadboldmathitalic
\loadboldgreek

\title{Multifractal structure of Ly$\alpha$ clouds: An example with the 
spectrum of QSO 0055--26}
\author[Vincenzo Carbone$^1$ and Sandra Savaglio$^{1,2}$]
       { Vincenzo Carbone$^1$ and Sandra Savaglio$^{1,2}$\\
        $^1$Dipartimento di Fisica, Universit\'a della Calabria, I--87036 Roges di 
Rende (CS) Italy\\
$^2$European Southern Observatory, K. Schwarzschild Stra$\beta$e 2, D--85748 Garching, Germany}
\date{Accepted 1996 February.
      Received...; 
      in original form 1995 December}
\pubyear{1994}

\begin{document}
\maketitle

\begin{abstract}

Ly$\alpha$ forests are usually associated with intergalactic gas clouds
intercepting the quasar sight--line.  Using as an example the Ly$\alpha$
forest of QSO 0055--26 ($z_{em} = 3.66$), we show that the probability of
observing a line at the velocity difference $\Delta v$ has an intermittent
structure in the redshift space.  On small scales (70 km~s$^{-1} \leq \Delta
v \leq 1580$ km~s$^{-1}$) the signature of intermittency appears as a
self--similar structure of spikes systematically visible at the same redshift
on all scales.  This behaviour can be interpreted as due to the presence of
clustering of the Ly$\alpha$ lines, which appear as singularities of the
density of the probability measure.  From a direct measurement of the
generalized dimensions, we show that the intermittency can be described by a
multifractal structure which is due to a Fourier phase correlation of the
signal.  The multifractal structure disappears at scales larger than $\Delta
v \simeq 1580$ km~s$^{-1}$.

\end{abstract}

\input epsf

\section{Introduction}

The numerous absorption lines observed in the quasar spectra, and commonly
referred as the Ly$\alpha$ forest, are usually ascribed to 
intergalactic gas clouds which intercept the quasar
sight--line (Sargent et al., 1980). In this sense Ly$\alpha$ lines represent 
good 
tracers of the intergalactic matter distribution. Up to now data analysis of 
Ly$\alpha$ forests is essentially based on the traditional statistical 
methods, like for example the two--point correlation function. However it 
would be quite interesting to perform an analysis where the traditional 
methods are replaced, or complemented, by the investigation of the scaling 
properties of the Ly$\alpha$ column densities. 
In fact, the two--point correlation function has failed to highlight 
without ambiguity the 
presence of clustering in the Ly$\alpha$ forest on low resolution data (Sargent 
et al. 1980; Bechtold et al. 1987; Webb \& Barcons 1991). High resolution 
observations (Chernomordik 1995; Cristiani et al. 1996) has revealed 
clustering up to scales of about 300 km~s$^{-1}$.

We attempt to apply a method where spurious effects due to the finite size of
the sample and to the poorly known column density distribution, can be
controlled. The aim is to unambiguously recognize the presence of clusters or
voids in the sample without losing information on their localization. In fact
gravitational clustering should manifest itself as structures localized in
redshift, detectable on all the dynamically interesting scales. The statistical
analysis which we would like to perform is based on the idea that clustering is
due to a Fourier phase correlation, which generates intermittency. The scaling
behaviour of a probability measure derived from the intermittent signal is
singular and can be described by the multifractal geometry. A quite similar
approach has recently been used also by Pando \& Fang (1995) who analyzed
the scaling behaviour in the structure of the QSOs Ly$\alpha$ lines by using the
wavelets decomposition analysis. Starting from the work by Frisch \& Parisi
(1985), multifractals have been invoked to describe many physical phenomena
where anomalous scaling laws are present. Examples include chaotic dynamical
systems, the rate of energy dissipation in fluid flows etc (Halsey et al., 
1986, and the reviews by Paladin \& Vulpiani, 1987; Meneveau \&
Sreenivasan, 1991, and references therein). Multifractal geometry in astronomy
has been used mainly to describe the mass distribution of galaxies (among 
others see for example Martinez et al., 1990; Jones et al., 1992;
Coleman \& Pietronero, 1992; 
Borgani et al., 1993, 1994; Martinez \& Coles, 1994), 
the apparent luminosity
field of galaxies (Garrido et al., 1996),
the cosmic
microwave background radiation (Pompilio et al., 1995) and the intermittency in
the Solar Wind turbulence (Carbone, 1993).  However the literature on the
subject has increased very rapidly in the last few 
years, and the multifractal analysis
is now a standard way to recover anomalous scaling laws, if any exist, from
random signals (Borgani, 1993; Vainshtein et al., 1994).

\section{Data analysis: The multifractal structure}

\begin{figure}
\caption{\label{fig0} Column densities of HI in the spectrum of
QSO 0055--26 vs.~$z$.}
\epsfxsize=9cm
\epsfysize=4.5cm
\epsffile{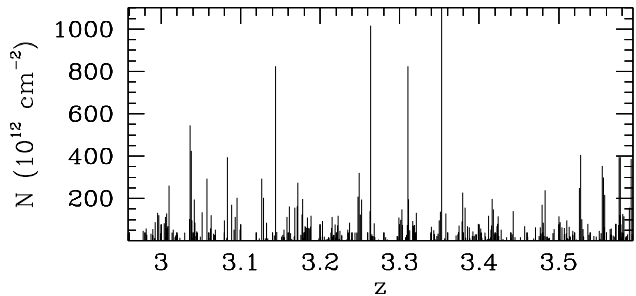}
\end{figure}

The aim of the present paper is to study the scaling behaviour of the HI
column 
densities $N(z)$ of Ly$\alpha$ clouds by using the absorption line list of 
quasars. 
As an example of the suitability of the method to capture the clustering due to 
phase correlations, we show the results relative to the
high resolution spectrum (FWHM $= 14$ km~s$^{-1}$) of the $z_{em} = 
3.66$ QSO 0055--26 (Cristiani et al. 1995). The observations in the
interval $4750 < \lambda < 
6300$ \AA~ cover a Ly$\alpha$ forest in the redshift range $2.958 < z < 
3.654$.
The Ly$\alpha$ lines have HI column densities 
in the interval $12.8 \leq \log_{10} N \leq 15.2$.
The data acquisition, reduction and 
analysis are described in Cristiani et al. (1995). The distribution
of the Ly$\alpha$ lines with column density $N$
along the redshift direction is shown in Figure \ref{fig0}.

\begin{figure}
\caption{\label{fig1} Behavior of the probability measure 
$P_i(\Delta v)$ 
as a function of the velocity $v$, for different scales $\Delta v$.}
\epsfxsize=9cm
\epsfysize=19cm
\epsffile{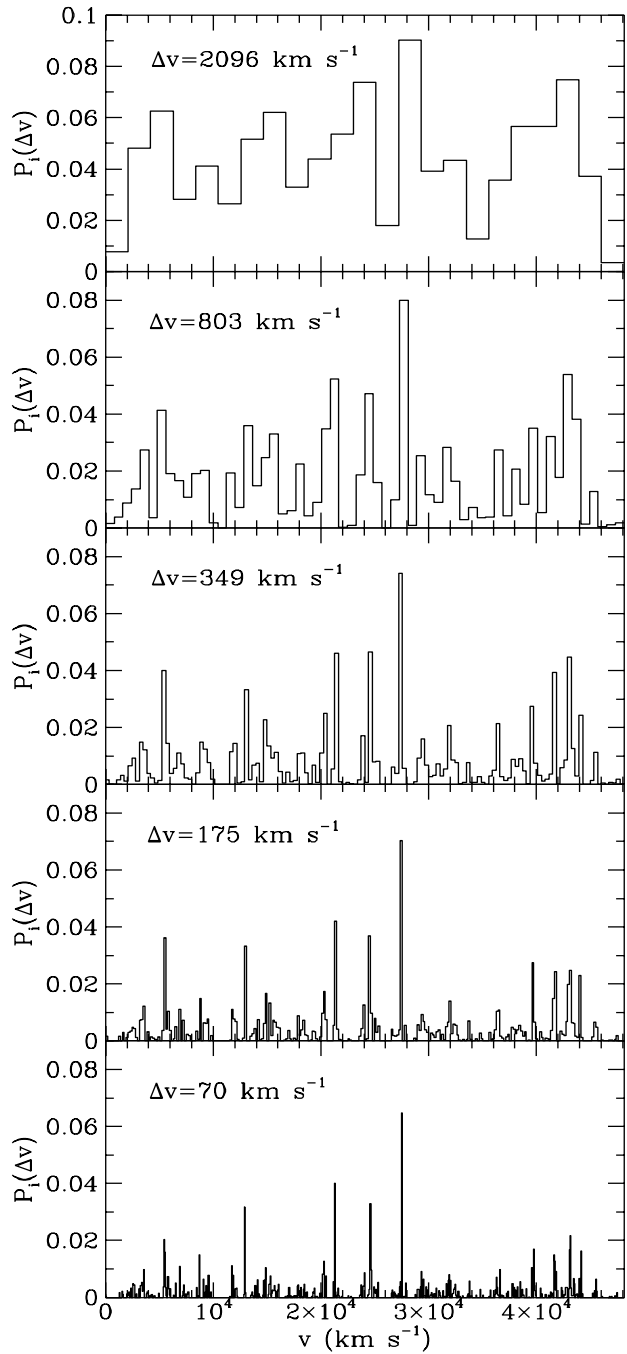}
\end{figure}

Two properties are evident from Figure \ref{fig0}: ({\it i}) $N(z)$
looks like a  stochastic variable; ({\it ii})
this variable has an intermittent nature.
Intermittency here is used in the same sense as it is used in fluid dynamics
for the dissipation of kinetic energy (Meneveau \& Sreenivasan,
1991), that is regions of large activity are
interspersed  
between those where the absorption line density is relatively
depressed.
This phenomenon, in the framework of galaxy distribution, has
been named ``heterotopic intermittency'' (Jones et al., 1992).
 As a reference the reader can note the
striking analogy between our Figure \ref{fig0} and the Figure 1 
of Meneveau \& Sreenivasan (1991) showing the typical time
evolution of the one--dimensional energy dissipation rate in turbulent
flows. This behaviour can be thought of as representing the near--singular
characteristics of the phenomenon. 
To see this, we make a scaling analysis of $N(z)$ in the velocity space,
introducing the velocity
difference $\Delta v$ between two redshifts $z_1$ and $z_2$
(Sargent et al., 1980)

\begin{equation}
{\Delta v \over c} = {(z_1+1)^2-(z_2+1)^2 \over (z_1+1)^2+(z_2+1)^2}~.
\label{deltav}
\end{equation}

\noindent
For each scale $\Delta v$ we define a probability measure by dividing the 
redshift range into disjoint subsets $\Omega_i$, from equation 
\ (\ref{deltav}). This measure $P_i(\Delta v)$ is defined as the total column 
density in the $i$--th subset characterized by a velocity separation 
$\Delta v$, normalized to the total column density in the spectrum. 
This can be related to the 
probability of occurrence of a certain amount of gas in the $i$-th box at a 
certain scale $\Delta v$. In Figure \ref{fig1} we show $P_i(\Delta v)$ 
for different scales, as  function of the velocity  $v$ in the redshift 
range $2.958 < z < 3.654$. As  can be noted the 
general behaviour of the measure on a given scale appears to be similar to that on 
another scale. What is interesting is that strong recurrent peaks with 
the same velocity are systematically detected on all scales $\Delta v$, 
indicating highly localized clouds with a strong scale--independent 
probability for their column density. These structures are now identified 
as the signature of clusters of Ly$\alpha$ absorption lines within the forest. 

To investigate the multifractal structure of the measure, we use the usual 
box--counting method (Halsey et al., 1986), by defining the 
generalized partition function

\begin{equation}
\chi^{(q)}(\Delta v) = \sum_i \left[P_i(\Delta v)\right]^q~,
\end{equation}

\noindent
where the sum is extended to all the subsets $\Omega_i$ at a given scale 
$\Delta v$. High values of $q$ enhance the strongest singularities (say the 
most intense clusters), while small values of $q$ evidence the regular
regions. It is evident that negative values of $q$ in the
partition function emphasizes the low density regions where the
probability  $P_i(\Delta v)$ is low.
The information relative to the multifractal structure can be recognized by 
calculating the generalized R\'enyi dimensions $D_q$ from the scaling law

\begin{equation}
\label{equ2}\chi^{(q)}(\Delta v) \sim \left[\Delta v\right]^{(q-1)D_q}
\end{equation}

\noindent
or the singularity spectrum  $f(\alpha)$ (Halsey et al., 1986).
This last quantity represents the dimension associated with each
singularity of strength $\alpha$ defined through $P_i(\Delta v) \sim 
(\Delta v)^{\alpha}$. The singularity spectrum is nothing but the
Legendre transform of $(q-1)D_q$, so that it can be recognized from
the measurement of $D_q$.
For a monofractal, $D_q$ is constant for each $q$. By contrast, if the 
intermittency can be described by multifractality we have $D_p < D_q$ for 
$p > q$. For example $D_0$ is the dimension of the support of the 
measure, which indicates the degree of fillness of the observed
redshift range. 
The absorption lines are concentrated asymptotically on a set of dimension 
$D_1$, while $D_2$ is the correlation dimension (Grassberger 
\& Procaccia, 1983). 

Some care must be used in performing the analysis just described, in fact
some effects can induce evidence of spurious multifractal structures. This is 
due to the lack of 
statistics, for high $q$, in samples with a finite number of points. To check 
 the genuine 
multifractal structure of the absorption lines, we examine the properties of 
their Fourier phases (see for example Pompilio et al., 1995). The 
idea comes from the fact that intermittency, and thus localized clusters, are 
due to Fourier phase correlations of the signal. For this reason 
we built up a set of $2000$ ``fake'' sequences of column densities $N_f(z)$ 
obtained from the 
true signal $N(z)$ by a process of phase randomization. In other words we 
have calculated the Fourier coefficients for the observed sequence $N(z)$, then the 
Fourier amplitudes of these coefficients are used to generate a new sequence of 
coefficients 
by using the same amplitudes but random phases. Finally, by the inverse Fourier 
transform, we obtain a given sequence $N_f(z)$. Each fake 
sequence, which has the same spectrum as the observed one, is 
then used to perform the same multifractal analysis we 
have described so far and compared  with that obtained with 
the true line distribution.

\begin{figure}
\caption{\label{fig2} Values of $\log_{10} \chi^{(2)}(\Delta v)$ vs. 
$\log_{10} \Delta v$. 
The straight lines represent the fit in the small and large scales.}
\epsfxsize=8cm
\epsfysize=7.2cm
\epsffile{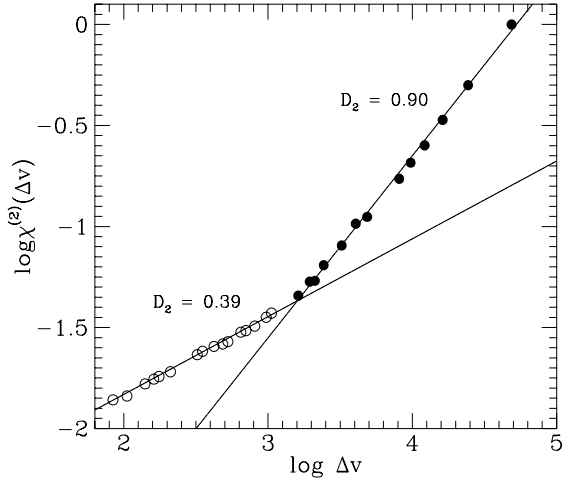}
\end{figure}

In Figure \ref{fig2} we show the values of $\log_{10} \chi^{(2)}(\Delta v)$ vs. 
$\log_{10} \Delta v$. The interesting behaviour is the fact that there exist two
distinct ranges where a linear relation can be found, and in these ranges we
obtain two different values of $D_2$ from equation \ (\ref{equ2}). This
happens for all the values of $q$. The presence of two distinct linear ranges 
is due to a ``lacunarity effect''. In fact the 
absorption lines are very localized in redshift, the error for the 
determination of a line being actually of the order of $10^{-5}$ 
(Cristiani et al., 1995). As a consequence, large gaps are present 
in between the absorption lines, and the support of the probability measure 
forms a random Cantor dust. The same effect is visible in the mass distribution 
of galaxies (Martinez \& Coles, 1994), and is obviously visible for each 
sequence $N_f(z)$. The separation of the scaling behaviour, for all our 
sequences, happens at about $\Delta v_{\ast} \simeq 1580$ km~s$^{-1}$, a scale which 
physically separates two different regimes. In the following we will distinguish 
between the large--scales where $\Delta v > \Delta v_{\ast}$ and the 
small--scales where $\Delta v < \Delta v_{\ast}$. 

\begin{table}
\caption{The values of the generalized dimensions for both the small--scales
(70 km~s$^{-1} \leq \Delta v \leq 1580$ km~s$^{-1}$) and 
for large--scales ($\Delta v > 1580$ km~s$^{-1}$). $D_q$ and $D_q^f$ represent 
the generalized dimensions obtained by using respectively the 
spectrum of QSO 0055--26 and the average value obtained from the set of 2000 
fake sequences.}
\begin{center}
\begin{tabular}{ccccc}
\hline\hline&&&&\\[-5pt]
\multicolumn{1}{c}{}
& \multicolumn{2}{c}{small--scales}
& \multicolumn{2}{c}{large--scales} \\
q & $D_q$  & $D_q^f$  & $D_q$ & $D_q^f$ \\
[2pt]\hline&&&&\\[-8pt]
0 & $0.72\pm0.02$ & $0.72\pm0.02$ & $1.00\pm0.00$ & $1.00\pm0.00$ \\
1 & $0.50\pm0.01$ & $0.65\pm0.03$ & $0.94\pm0.01$ & $0.93\pm0.02$ \\
2 & $0.39\pm0.02$ & $0.60\pm0.05$ & $0.90\pm0.01$ & $0.90\pm0.03$ \\
3 & $0.27\pm0.03$ & $0.55\pm0.07$ & $0.88\pm0.03$ & $0.88\pm0.03$ \\
4 & $0.19\pm0.05$ & $0.49\pm0.10$ & $0.87\pm0.05$ & $0.86\pm0.06$ \\
[2pt]\hline\end{tabular}\end{center}
\end{table} 

In Table 1 we report the values of the generalized dimensions $D_q$ and $D_q^f$
(calculated as the mean of the 2000 fake values), obtained through equation 
\ (\ref{equ2}).  Some features must be discussed, the most evident being the
remarkable difference which we found between the small--scales and the
large--scales.  By looking at the small--scales we can see that $D_q$ is not
constant, that is it behaves as a nonlinear function of $q$. This indicates the
presence of a multifractal structure. To see that the differences between the
various values of $D_q$ are meaningful, we compare these values with
$D_q^f$. As can be seen $D_q < D_q^f$ for each $q$, which allows us to 
conclude that a multifractal structure, due to phase correlation, underlies the 
observed sequence. The decrease of
$D_q^f$ as $q$ increases, indicates a spurious multifractality not due to phase
correlations which would tend to disappear for richer samples of lines. However 
the genuine multifractality of $N(z)$, due to phase correlations, is evident. As
regards the large--scale behaviour we can see that the  generalized
dimensions obtained both from the true and from the 
fake samples show the same behaviour,
with the same values.  This should be interpreted as the absence of
intermittency and the lack of a true multifractal structure beyond the scale
$\Delta v_{\ast} \simeq 1580$ km~s$^{-1}$.  Even in this case the residual spurious
multifractality could be due to the limited number of data points in our sample.

\section{Discussion}

In order to better understand 
the structure of the Universe and to find, if any, the
scale at which it  becomes homogeneous, a large sample of objects with
a large spatial distribution is required. In this respect, the QSO absorption
spectra available to the scientific community represent a valuable test, being
distributed in a much more extended region (which can reach a space coverage of
several hundred $h^{-1}$ Mpc in comoving distances) with respect to the
observed galaxy distribution, and forming a more unbiased sample because the
absorbing objects are not selected according to luminosity criteria.

In the present paper we performed a statistical analysis of the Ly$\alpha$
absorption lines based on the multifractal formalism. The main motivation to do
this analysis is the fact that the behaviour of the column density $N(z)$ with
redshift shows a highly intermittent structure where singularities can be
detected. Since in the multifractal formalism different regions at a given
scale are weighted in different ways, as $q$ is increased the regions where the
strongest column densities $N(z)$ lie are weighted differently from the regions
where $N(z)$ are small or absent. Through a scaling analysis we are then able
to detect the signature of clustering due to gravitational effects. The
presence of a multifractal structure in the behaviour of $N(z)$ should indicate
the presence of a hierarchy of intergalactic Ly$\alpha$ clouds, probably
generated through contraction effects (M\"ucket et al.,1995). We found this
structure at smaller scales, say in the range 70 km~s$^{-1} \leq \Delta v \leq 1580$ 
km~s$^{-1}$
(corresponding to comoving 
distances of about $0.5 \leq r \leq 7.6\ h^{-1}$ Mpc). At scales larger than 
$7.6\ h^{-1}$ Mpc the multifractal structure disappears and the structure is
homogeneous. It is 
interesting to note that the maximum scale of clustering we obtain is larger than that 
obtained from preliminary results with the two--point correlation 
function (Chernomordik 1995; Cristiani et al., 1996), which in fact found 
clusters up to a scale $\Delta v \simeq 300$ km~s$^{-1}$. 

High resolution spectroscopy of quasars has recently received a
boost by observations with the HIRES spectrograph of the Keck telescope
(Fan \& Tytler 1994; Hu et al. 1995). It is important to see whether the
scaling behaviour is also present in other available spectra. 
Preliminary results have shown that the scaling
properties persists in different quasar sight--lines. The generalized dimension
for small scales is consistent with what found for QSO 0055--26, but 
the characteristic scale $\Delta v_*$  shows
a redshift dependence. A detailed discussion will be presented
in a forthcoming paper (Savaglio \& Carbone, 1996).

A comparison of our results with those obtained from the same analysis when
applied to galaxy catalogues (CfA1, CfA2 and QDOT redshift survey) and to the
Abell and ACO catalogs of galaxy clusters, should allows us to obtain a more
unified picture of the mass distribution. In fact from the CfA survey one
obtains significantly smaller values for the generalized dimensions, say in a
three-dimensional embedding space $D_0^{(3)} \simeq 2.1$, and
$D_2^{(3)} \simeq 1.3$ for a
wide range of scales (the relation between the one--dimensional cut and the 
$d$-dimensional cut is simply $D_q^{(d)}=(d-1)+D_q^{(1)}$). 
On the contrary the Abell and
ACO clusters (Borgani et al., 1994) show a scale--invariant multifractal
structure only in a limited range of scales, for Abell the range is 
$15-60\ h^{-1}$ Mpc, while for ACO it extends to smaller scales. In both cases
$D_2^{(3)} \simeq 2.2$, and the picture of a pure scale invariant fractal structure
extending to larger distances is disproved by these analyses. Finally the 
multifractal scaling properties of the QDOT redshift survey (Martinez \& Coles, 
1994) extend over two well defined scaling ranges, for $10-50\ h^{-1}$ 
Mpc and for $1-10\ h^{-1}$ Mpc. In both cases $D_0^{(3)} \simeq 2.9$
while $D_2^{(3)} 
\simeq 2.77$ at large scales and $D_2^{(3)} 
\simeq 2.25$ at small scales. These last 
results are not really different from those obtained in our case. As
regards the 
CfA surveys and the galaxy clusters, our results would highlight the fact that 
galaxies and the Ly$\alpha$ cloud have different behaviour in space. This would 
be natural in a universe where the baryonic dark matter is distributed in a 
homogeneous way, as most of the cosmological models imply and the COBE results 
on the cosmic microwave background radiation show.

So far we have calculated the values of $D_q$ for positive values of $q$, 
because this part of the spectrum emphasizes the regions characterized by 
high densities of the measure (clusters). Even if there is a lack of 
statistical significance, due to the fact that the statistics of rare events 
requires very long data sets, we found two scaling regimes with multifractal 
scaling for $\Delta v < \Delta v_{\ast}$. By using the clustering paradigm 
(Martinez et al., 1990; Martinez \& Coles, 1994), this implies the 
presence of clusters localized at these scales. On the contrary the part of the spectrum $D_q$ with negative values of $q$ characterizes the low density 
regions. In this way we should be able to recognize the  presence, if any, of 
voids. Unfortunately (Borgani et al., 1993) discreteness effects and lack 
of statistics heavily affect the evaluation of $D_q$ for negative $q$, and 
this effect is stronger than the lack of statistics which affects the
positive part of the spectrum.

\section*{Acknowledgments}
We are grateful to A. Provenzale for useful discussions. It is a
pleasure to thank F. Malara and P. Veltri for their interest on this work.


\begin{thebibliography}{99}
\bibitem{b1}
Bechtold, J., Green, R.F., \& York, D. 1987, ApJ, 312, 50

\bibitem{b1}
Borgani, S. 1993, MNRAS, 260, 537

\bibitem{b1}
Borgani, S., Murante, G., Provenzale, A., \& Valdarnini, R. 1993, Phys. Rev. E, 47, 3879

\bibitem{b1}
Borgani, S., Martinez, V.J., P\`erez, M.A., \& Valdarnini, R. 1994, ApJ, 435, 37

\bibitem{b1}
Carbone, V. 1993, Phys. Rev. Lett., 71, 1546

\bibitem{b1}
Chernomordik, V.V. 1995, ApJ, 440, 431

\bibitem{b1}
Coleman, P.H., \& Pietronero, L. 1992, Phys. Rep., 213, 311


\bibitem{b1}
Cristiani, S., D'Odorico, S., D'Odorico, V., Fontana, A., Giallongo, E., \& Savaglio, S. 1996, 
MNRAS, submitted

\bibitem{b1}
Cristiani, S., D'Odorico, S., Fontana, A., Giallongo, E., \& Savaglio, S. 1995, 
MNRAS, 273, 1016

\bibitem{b1}
Garrido, P., Lovejoy, S., Schertzer, D. 1996, Physica A, {\it in
press}

\bibitem{b1}
Frisch, U., \& Parisi, G. 1985, in {\it Turbulence and Predictability in 
Geophysical Fluid Dynamics}, edited by M. Gil, R. Benzi and G. Parisi 
(North--Holland, Amsterdam), pag. 84

\bibitem{b1}
Grassberger, P., \& Procaccia, I. 1983, Phys. Rev. A, 28, 259

\bibitem{b1}
Halsey, T.C., Jensen, M.H., Kadanoff, J.M.H., Procaccia, L.P., \& Shraiman, B.I. 1986, 
Phys. Rev. A, 33, 1141

\bibitem{b1}
Jones, B.J.T., Coles, P., Martinez, V.J. 1992, MNRAS, 259, 146

\bibitem{b1}
Martinez, V.J., \& Coles P. 1994, ApJ, 437, 550

\bibitem{b1}
Martinez, V.J., Jones, B.J.T., Dominguez--Tenreiro, R., \& van de Weygaert R. 
1990, ApJ, 357, 50

\bibitem{b1}
Meneveau \& Sreenivasan K.R. 1991, J. Fluid Mech., 224, 429

\bibitem{b1}
M\"ucket, J.P., Petitjean, P., Kates, R.E., \& Riediger, R. 1996, {\it
in press}

\bibitem{b1}
Paladin, G., \& Vulpiani, A. 1987, Phys. Rep., 156, 147

\bibitem{b1}
Pando, J., \& Fang L.-Z. 1996, ApJ, 459, 1
preprint

\bibitem{b1}
Pompilio, M.P., Bouchet F.R., Murante, G., \& Provenzale, A. 1995, ApJ, 449, 1

\bibitem{b1}
Sargent, W.L.W., Young. P.J., Boksenberg, A., \& Tytler, D. 1980, ApJS, 42 41

\bibitem{b1}
Savaglio, S., Carbone, V. 1996, {\it in preparation}

\bibitem{b1}
Vainshtein, S.I., Sreenivasan, K.R., Pierrehumbert, R.T., Kashyap, V., \& 
Juneya, A. 1994, Phys. Rev. E, 50, 1823

\bibitem{b1}
Webb, J.K. 1987, in  {\it IAU Symposium 124, Observational Cosmology}, ed. 
A. Hewett, G. Burbidge, \& L.Z. Fang (Dordrecht: Reidel), 803

\bibitem{b1}
Webb, J.K., \& Barcons, X. 1991 MNRAS, 250, 270

\end{thebibliography}
\end{document}